\documentclass[aps,prl,twocolumn,superscriptaddress,groupedaddress]{revtex4}  
\usepackage{graphicx}  
\usepackage{dcolumn}   
\usepackage{bm}        
\usepackage{amssymb}   
\usepackage{amsmath}
\usepackage{color}
\usepackage[colorlinks=true, pdfstartview=FitV, linkcolor=blue, citecolor=blue, urlcolor=blue]{hyperref} 
\usepackage{epstopdf}

\usepackage{appendix} 

\newcommand{\ket}[1]{\left| #1 \right>} 
\newcommand{\bra}[1]{\left< #1 \right|} 

\definecolor{orange}{RGB}{255,127,0}

\begin{document}

\preprint{}

\title{Superlattice switching from parametric instabilities in a driven-dissipative BEC in a cavity}
\author{Paolo Molignini}
\thanks{Authors contributed equally.}
\affiliation{Institute for Theoretical Physics, ETH Zurich, 8093 Zurich, Switzerland}
\author{Luca Papariello}
\thanks{Authors contributed equally.}
\affiliation{Institute for Theoretical Physics, ETH Zurich, 8093 Zurich, Switzerland}
\author{Axel U. J. Lode}
\affiliation{Wolfgang Pauli Institute c/o Faculty of Mathematics,
University of Vienna, Oskar-Morgenstern Platz 1, 1090 Vienna, Austria}
\affiliation{Department of Physics, University of Basel, 4056 Basel, Switzerland}
\affiliation{Vienna Center for Quantum Science and Technology,
Atominstitut, TU Wien, Stadionallee 2, 1020 Vienna, Austria}
\author{R. Chitra}
\affiliation{Institute for Theoretical Physics, ETH Zurich, 8093 Zurich, Switzerland}
%

\begin{abstract}
We numerically obtain the full time-evolution of a parametrically-driven dissipative Bose-Einstein
condensate in an optical cavity and investigate the implications of driving for the phase diagram.
Beyond the normal and superradiant phases, a third nonequilibrium phase emerges as a many-body
parametric resonance. This dynamical normal phase switches between two symmetry-broken
superradiant configurations. The switching implies a breakdown of the system's mapping to
the Dicke model. Unlike the other phases, the dynamical normal phase shows features of nonintegrability and thermalization.   
\end{abstract}


\maketitle


\textit{Introduction} -- 
Quantum light-matter systems  present an ideal platform to study the confluence of  many-body physics and  time-periodic modulations~\cite{Ritter2009, PRLPurdy}. 
High-frequency  modulation  is an established versatile tool to experimentally engineer a wide array of static hamiltonians~\cite{eckardt, PRL_Meinert2016, goldmanPRX}. 
Near-resonant driving, on the other hand, provides a means of combining unique phenomena like parametric resonance~\cite{Landau_Lifshitz}, dynamical 
localization~\cite{Casati, Fishman} and collective many-body physics.   
Though driven interacting systems tend to heat up~\cite{Lazarides, Alessio, Ponte}, the interplay between periodic driving and cooperative effects offers the exciting possibility of realizing  
exotic prethermalized steady states.

Superradiant phases are the   quintessential  example of collective behavior in light-matter systems~\cite{Keeling}.
The  classic model describing this is the Dicke model~\cite{Dicke:1954, Hepp:1973, PRLdomokos, Dimer}, where a  quantum cavity mode collectively couples to independent 
two-level atoms~\cite{Hepp:1973, PRLdomokos, Dimer}.   
This physics was recently realized experimentally  in a weakly interacting cold bosonic gas coupled to a high finesse optical 
cavity~\cite{Baumann:2010, Baumann:2011,  Klinder17032015, PRLBakhtiari}, where superradiance manifests itself via the spontaneous formation of a lattice
supersolid~\cite{Brennecke:2013}.  
Easy implementations of parametric modulations in  a wide range of frequencies make it  the perfect  realistic system  in which to study the influence of drive, 
interaction as well as dissipation. 
This is particularly interesting as parametric  modulation of the light-matter coupling  in the Dicke model (DM) was shown to generate a parametric 
instability~\cite{bastidasPRL, Chitra:2015} to an intriguing new phase of matter, termed the dynamical normal phase (D-NP)~\cite{Chitra:2015}.

\begin{figure*}
\centering
\includegraphics[width=\textwidth]{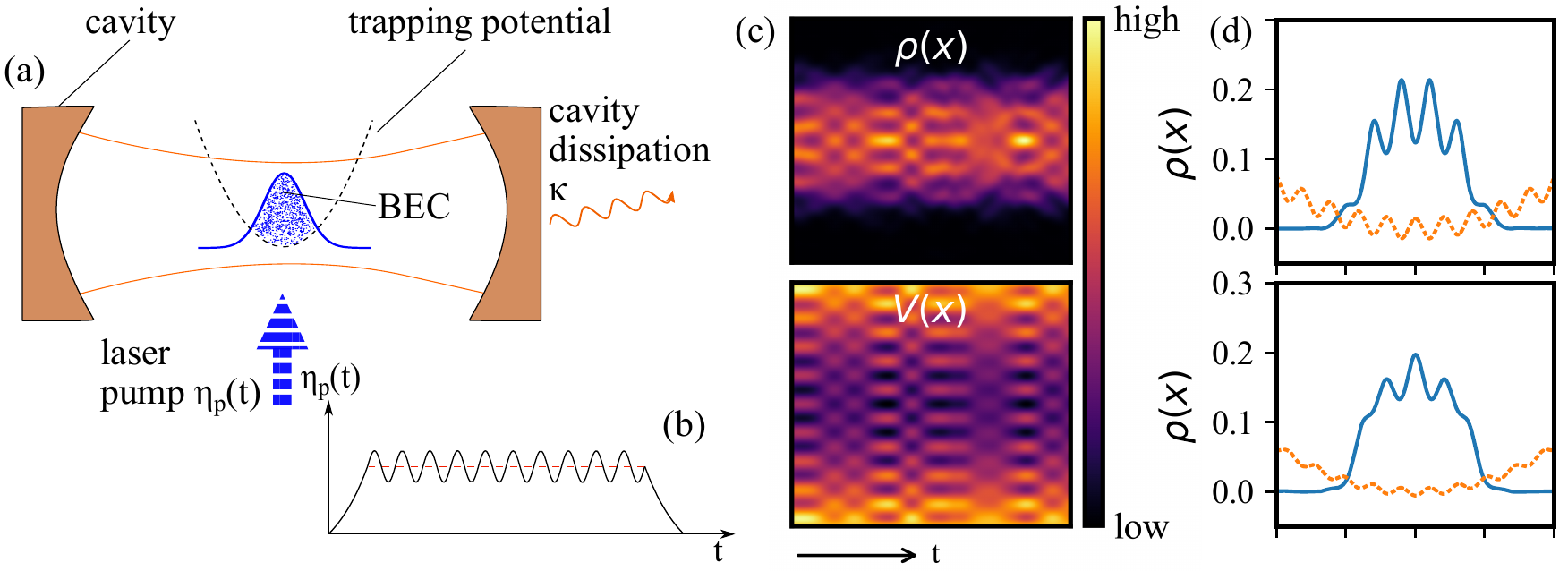}
\caption{ (a) A BEC in a transversely pumped dissipative cavity, subject to an external harmonic potential of frequency $\omega_x$.  
(b) Laser pump profile $\eta_p(t)$, comprising a ramp-up, a sinusoidally modulated plateau and a 
ramp-down. (c) Time-evolution of the density $\rho(x,t)$ (upper panel) and the self-consistent one-body potential $V(x,t)$ (lower panel) for a time-interval pertaining to the 
plateau for the D-NP. 
(d) Snapshots of the atomic density $\rho (x, t)$ (solid blue line) and one-body potential $V (x, t)$ (dashed orange line) seen by the atoms in the modulated plateau region.
In the upper panel, the BEC is mostly localized on the even sites of the periodic lattice. In the lower panel, the BEC is mostly localized on the odd sites. 
The drive causes atoms of the BEC oscillate in time between even and odd sites of the periodic lattice in the D-NP.  }
\label{figure-setup}
\end{figure*}

In this Letter, we obtain the phase diagram of a Bose-Einstein condensate (BEC) in a dissipative optical cavity with parametrically modulated atom-cavity coupling 
[see Fig.~\ref{figure-setup} (a) and (b)].  Many-body parametric resonance occurs in this system resulting in an emergent oscillatory phase of matter. 
In this phase, the drive facilitates the dynamical switching of the system between the two symmetry-broken ordered configurations permitted by the undriven Hamiltonian. 
This switching is, however, explicitly forbidden in the static case. Contrary to standard expectations, this interacting driven-dissipative system displays heating characteristics which depend
on its phase, further enriching its physics.

\textit{Model} -- 
The BEC comprises $N$ interacting atoms and is dispersively coupled to a high-finesse optical cavity with a single mode of frequency 
$\omega_c$, Fig.~\ref{figure-setup}(a). 
The atoms of the BEC have a transition frequency $\omega_a$ and are coherently driven by a transverse pump laser of frequency $\omega_p$. 
In the dispersive-coupling regime, if the cavity and atoms are strongly detuned in the rotating frame, i.e., $\Delta_c = \omega_p - \omega_c \gg \omega_a$ and 
$\Delta_a = \omega_p - \omega_a \gg \omega_a$, one can adiabatically eliminate the excited atomic levels~\cite{Nagy:2008} to obtain the following effective Hamiltonian 
for the coupled BEC-cavity system in the rotating frame:
\begin{widetext}
\begin{align}
\mathcal{H}_{\textrm{BEC}} 
&  = 
\int \mathrm{d}^3 r \, \hat{\Psi}^{\dagger}(\mathbf{r}, t) 
\left\{ 
-\frac{\hbar^2}{2m} \nabla^2 + V_{\textrm{trap}}(\mathbf{r}) + \frac{U}{2}  \hat{\Psi}^{\dagger}(\mathbf{r}, t)  \hat{\Psi}(\mathbf{r}, t) 
\right\} 
\hat{\Psi}(\mathbf{r}, t) - \hbar \Delta_c \hat{a}^{\dagger} \hat{a}   \nonumber \\
& \quad + \int \mathrm{d}^3 r \, \hat{\Psi}^{\dagger}(\mathbf{r}, t) \left\{ 
\frac{\hbar}{\Delta_a} \left[ h^2(\mathbf{r}, t) + g^2(\mathbf{r}) \hat{a}^{\dagger}\hat{a} + h(\mathbf{r}, t) g(\mathbf{r}) ( \hat{a} + \hat{a}^{\dagger}) \right] \right\} 
\hat{\Psi}(\mathbf{r}, t) .
\label{ham:BEC}
\end{align}
\end{widetext}
The atoms in the BEC are described by bosonic field operators $\hat{\Psi}^{(\dagger)}(\mathbf{r}, t)$ while $\hat{a}^\dagger$ and $\hat{a}$ describe the cavity mode. 
All operators obey bosonic commutation relations.
For the sake of computational simplicity, in the following we restrict ourselves to the one-dimensional problem along the cavity axis $x$. 
The atoms are subjected to a harmonic trapping potential $V_{\textrm{trap}} (x) = m \omega_x^2 x^2 / 2$ and interact through short-range interactions with the strength 
$U = 4 \pi \hbar^2 a / m$ where $a$ is the $s$-wave scattering length and $m$ the mass of the atom~\cite{HuangSM, Olshanii, PethickBEC}.
 
The atoms are driven by a transverse pump field described by the mode-function $h(\mathbf{r}, t) = h(z, t) = \eta_p (t) \cos k z$ while the cavity mode function is 
$g(\mathbf{r}) = g(x) = g_0 \cos k x$, where $\eta_p$ is the pump rate, $k$ the wavelength of the light and $g_0$ the atom-cavity coupling.
The last two terms in the Hamiltonian describe the atom-cavity interaction. The $g^2(\mathbf{r})$ term arises directly from the cavity mode function while the 
$h(\mathbf{r}, t) g(\mathbf{r})$-term results from the interference between cavity and pump fields. 

For static pumps a mean-field analysis for large $N$ and $V_{\textrm{trap}} (\mathbf{r}) = 0$ using the Gross-Pitaevskii equation 
shows a $\mathbb{Z}_2$-symmetry breaking transition; as the pump power increases, the system goes from a normal phase (NP) with no photons in the cavity to a superradiant phase (SP), 
where the cavity field is a coherent state~\cite{Nagy:2008, Baumann:2010, PRLNagy2010}. In the SP, the atoms spontaneously self-organize into either an even or odd lattice 
structure with lattice spacing $\lambda= \frac{2\pi}{k}$~\cite{Baumann:2010}. 
The relevant order parameter  is $\Theta \equiv \bra{\psi} \cos kx \ket{\psi}$:  $\Theta = 0$ in the NP and $\Theta \ne  0$ in the SP. $\Theta$  essentially  counts the population 
imbalance between odd and even  lattice sites in the  SP. The same physics is well-described by a mapping to the 
DM Hamiltonian which assumes that only the lowest $\pm k$-modes of the atoms are populated~\cite{PRLNagy2010,Baumann:2010}.

In this Letter, we study the parametrically modulated system described by the Hamiltonian \eqref{ham:BEC}.  
The time-dependent pump amplitude,
\begin{equation}
\eta_p(t) = \eta_p^0 \left( 1 + \alpha \sin \left( 2\pi t /T \right) \right), \label{eq:pump}
\end{equation}
will lead to a new phase of matter. We explicitly include the cavity dissipation.
We consider system parameters which describe the experimental system studied in Refs.~\cite{Baumann:2010, Baumann:2011} (see also~\cite{supmat}).
For the considered system, the cavity dynamics follows that of the atoms closely, because the cavity detuning $\Delta_c$ is much larger than the other energy scales. 
In this adiabatic limit, it is reasonable to replace $\hat{a}$ by its expectation value $\langle \hat{a} \rangle \equiv a$ in the dynamics for the atoms. 
This cavity order parameter obeys a dissipative equation of motion with a rate $\kappa$~\cite{supmat}. The resulting dynamical evolution of the atoms is then 
studied using the Multi-Configurational Time-Dependent Hartree method for indistinguishable particles (MCTDH-X)~\cite{ultracold, exact_F, SMCTDHB, MCTDHX}, 
see Ref.~\cite{supmat} for details. This method has been very successful in describing the dynamics of host bosonic 
systems~\cite{PRLAxel, MCTDHB_init, MCTDHB_PRL}. Here, for the first time, we apply the method to a periodically driven many-body system coupled to an 
optical cavity.

\begin{figure}[t]
\includegraphics[width=\columnwidth]{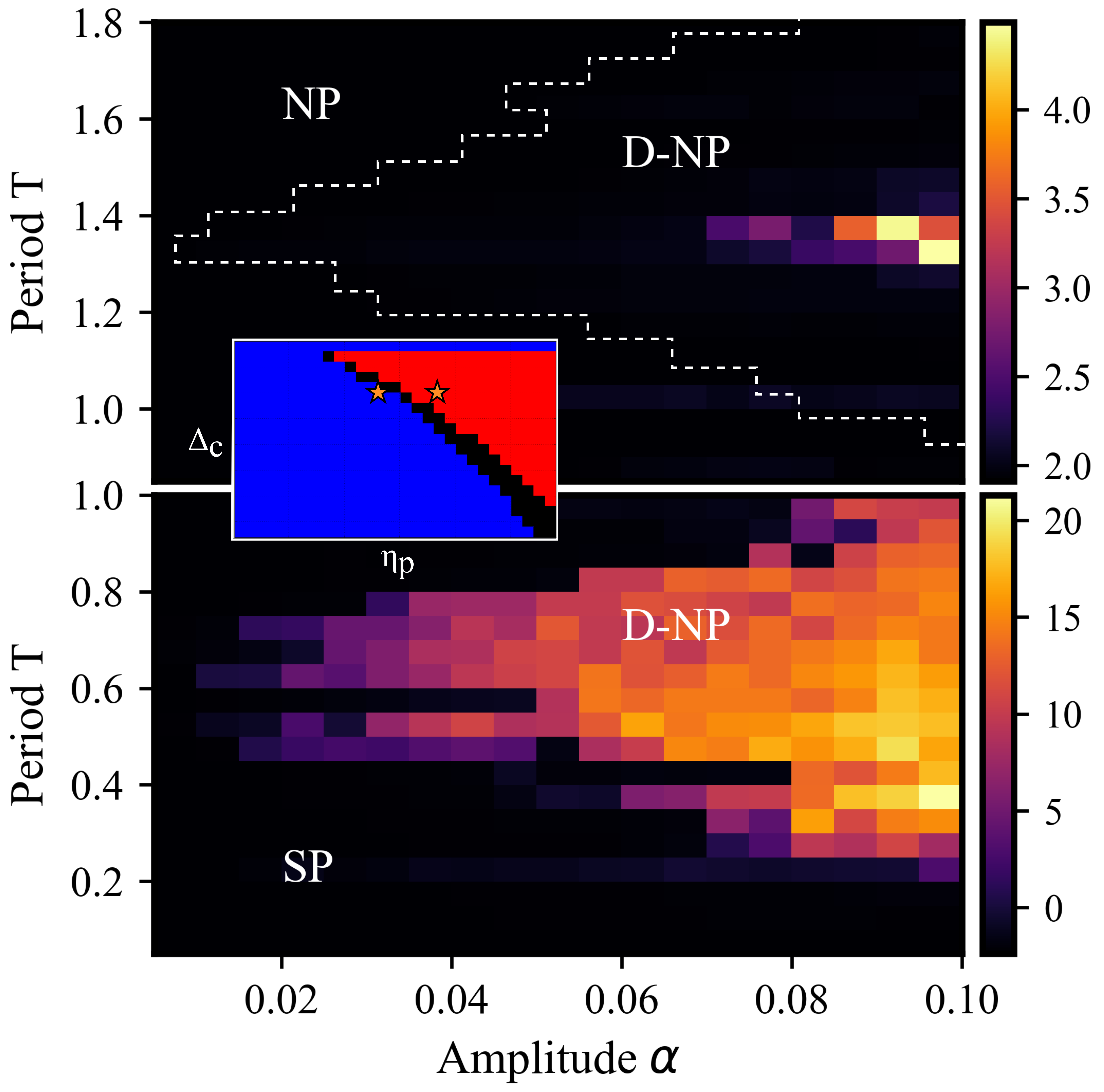}
\caption{Phase diagrams of a parametrically driven BEC in a cavity for $N=1000$ bosons. They are superimposed on the average heating of the system as a function of modulation amplitude 
$\alpha$ and period $T$. 
The inset shows the undriven phase diagram. The parameters for the driven cases in the NP ($\eta_p^{0,\textrm{NP}} = 2\pi \cdot 3.28 \: \text{kHz} , \Delta_c=- 2\pi \cdot 10.08 \: \text{MHz}$; phase 
diagram in the upper panel) and in the SP ($\eta_p^{0,\textrm{SP}} = 2\pi \cdot 4.79 \:  \text{kHz}, \Delta_c=- 2\pi \cdot 10.08 \: \text{MHz}$; phase diagram in the lower panel) are marked as stars 
in the inset.
In the upper panel the BEC is driven from the NP and the dashed line delineates the NP and the D-NP.  
In the lower panel the condensate is driven from the SP and the black region corresponds to the SP while the colored region corresponds to the D-NP. 
In both diagrams the Arnold tongue corresponds to the first resonance ($n=1$) with a period of $T_1 \approx 1.3$ (NP) and $T_1 \approx 0.65$ (SP). 
In both plots, the black regions indicate no heating while the coloured tiles indicate heating.  
All quantities shown are dimensionless, see~\cite{supmat}. }
\label{PDs}
\end{figure}

\textit{Results} -- 
As a benchmark, we reproduce the undriven $(\eta_p$, $\Delta_c$)-phase diagram obtained in~\cite{Baumann:2010}.
We use the time-dependent transverse pump protocol illustrated in Fig.~\ref{figure-setup}(b) with the modulation amplitude $\alpha=0$ in Eq.~\eqref{eq:pump}. 
At time $t=0$, the cavity is decoupled from the BEC in the trap. 
As  $\eta_p(t)$ is ramped up, and approaches a constant, $\eta_p^0$, in the plateau we obtain either the NP where 
$\Theta=0$, or the SP where $\Theta \ne 0$ (see inset in Fig.~\ref{PDs}). 
For $N=1000$, converged results are obtained using $M=1$ orbital in the MCTDH-X approach. Contributions from $M>1$ orbitals are negligible.
For this $M=1$ case, MCTDH-X corresponds to the mean-field solution obtained using the Gross-Pitaevskii equation~\cite{PethickBEC}.
We verified that our results recover the scaling invariance under $N \to N'$ provided  $g_0 \to \frac{N}{N'} g_0$, $U \to \frac{N}{N'} U$ and $\eta_p^0 \to \sqrt{\frac{N}{N'}} \eta_p^0$~\cite{Nagy:2008}.  

We now discuss a non-zero modulation $\alpha$ and assess the nature of the parametrically driven system when it is driven starting either from the NP or from the SP.
For a fixed detuning $\Delta_c=- 2\pi \cdot 10.08 \: \text{MHz}$, we select two representative points close to the SP-NP phase boundary of the static pump simulations (see inset in Fig.~\ref{PDs}). 
Note that $\alpha$ is chosen to be small enough, so that the instantaneous $\eta_p(t)$ never  crosses the  static phase boundaries.  In the spirit of standard parametric driving,
the modulating frequency $\omega\equiv 2\pi/T$ is  chosen to be close to twice the gap to the lowest polaritonic excitation in the system. 
The polaritonic gap is  determined by mapping the driven BEC-cavity to the DM~\cite{supmat}. 
We obtain two polaritonic  modes for each phase~\cite{Emary}: a very high energy branch, $\epsilon_{+}^{\textrm{NP/SP}}$ proportional to $\Delta_c$, and a low energy branch
$\epsilon_{-}^{\textrm{NP/SP}}$ proportional to the atomic recoil energy $E_r = \frac{\hbar^2 k^2}{m}$,
\begin{align}
(\epsilon_{-}^{\textrm{NP}})^2 & \approx E_r^2 \left( 1 - (\eta_p^0/\eta_{p, c})^2 \right) , \\
(\epsilon_{-}^{\textrm{SP}})^2 & \approx E_r^2 \left( (\eta_p^0/\eta_{p, c})^4 - 1 \right) .
\end{align}
The  energy of the lower polaritonic branch goes to zero at the QPT  where $\eta_p \to \eta_{p, c}$ (see  inset in Fig.~\ref{PDs}).  
To study the impact of parametric driving on the atoms, we simulate the full time-evolution of the system as a function of $\omega$ starting from both NP and SP. 

Our results for the phase diagram of the modulated BEC-cavity system as a function of the drive amplitude $\alpha$ and period $T$ are summarized in Fig.~{\ref{PDs}, obtained by evaluating the order 
parameter $\Theta$. In Fig.~{\ref{PDs} we have also superimposed a color plot of the time averaged energy profile. In the upper panel the phase boundary is indicated with a white dotted line, 
while for the lower panel the transition from low to high energy zones coincides with the phase boundary obtained from $\Theta$.
Both phase diagrams show the emergence of a many-body parametric resonance: the static phases display parametric instability lobes -- reminiscent of  Arnold lobes for Mathieu oscillators -- 
for certain resonant values of $T$~\cite{HillsEq, mclachlan}. 
This is related to the fact that in the DM, polaritonic excitations are effectively described by the physics of two coupled parametric oscillators~\cite{supmat}. 
As the parameters $(T,\alpha)$ are varied, the underlying undriven NP/SP become unstable and the system transitions to a new phase which we term the D-NP~\cite{Chitra:2015}. 
In this phase,  the order parameter $\Theta(t)$ shows oscillatory behavior in time with zero mean (excluding trap contributions). 
 
The instability lobes seen in Fig.~\ref{PDs} differ greatly from the standard Arnold lobes for parametric oscillators described by the Mathieu equation. 
The periods around which the Arnold lobes of the BEC-cavity system are centered can be calculated as follows. In both phases, the dynamics is essentially governed by a classical 
Hill equation~\cite{HillsEq, supmat}
\begin{equation}
\ddot{x} + \gamma \dot{x} +  [\epsilon_-^{\textrm{NP/SP}}(t)]^2 x = 0 ,
\label{hill:eqn}
\end{equation}
where $\gamma$ is some effective damping and $ \epsilon_-^{\textrm{NP/SP}}(t)\equiv \epsilon_-^{\textrm{NP/SP}}(\eta_p (t))$.  
The parametric resonance condition is determined by $\epsilon_-^{\textrm{NP/SP}}(t=0) / \omega = n / 2$, $n \in \mathbb{N}_0$ whereas the structure of the instability lobes is determined 
by the detailed form of $(\epsilon_-^{\textrm{NP/SP}})$. The many-body resonance periods in our simulations are in good agreement with this simple resonance condition for $n=1$ when the 
system is driven starting both from the NP and SP. The resulting lobes are the first instability lobes and the complex lobe shape in the SP case is qualitatively captured by 
Eq.~\eqref{hill:eqn}. Higher values of  the cavity dissipation $\kappa$ were found to smoothen the shape of the D-NP lobe (cf. Ref.~\cite{supmat}).


Insights into the nature of the different phases can be gained by the analysis of the time-evolved density $\rho(x, t)$ and the effective one-body potential $V (x, t)$ seen by the atoms (see~\cite{supmat}). 
In the NP, the trapped BEC has a Gaussian profile, $\Theta(t) \approx 0$ and $\rho(x, t)$ shows minimal changes as a function of time.
In the SP, the atoms occupy the sites of the even or the odd lattice and $\Theta$ shows an oscillatory behavior with nonzero mean (cf. Fig.~\ref{energy-profiles} (b) light blue line). 
In the modulated plateau region, cf. Fig.~\ref{figure-setup}, the atoms 
remain in the lattice configuration chosen by the atoms before the pump modulation was turned on.
In the D-NP, however, the atoms and their potential systematically oscillate between the even and the odd lattice configurations [see Fig.~\ref{figure-setup}(c) and (d)], hinting at a complex dynamical 
particle reconfiguration. As expected from the general solutions of Mathieu-like equations~\cite{HillsEq, mclachlan}, both the density and the one-body potential oscillate in time not at the 
underlying driving frequency [see Fig.~\ref{figure-setup}(c)], but rather aperiodically.
We see that the lattice contribution to the effective potential seen by the atoms goes to zero at the point where the atoms transition between the even and the odd lattice.


We now discuss the stability of the different phases to heating, which is endemic to periodically driven interacting systems.  
Figs.~\ref{energy-profiles}(a) and (b) show $\Theta$ and Figs.~\ref{energy-profiles}(c) and (d) show the time-evolution of the energy per particle  in the various phases. 
In the NP and SP, the energies oscillate (aperiodically) in time, but their time averages stay constant [see Fig.~\ref{energy-profiles}(c) and (d)].  
This suggests that the BEC-cavity system -- even though it is an interacting system -- does not absorb sufficient energy from the drive to counteract the dissipation. Therefore, it does not heat up 
in the parametrically driven NP and SP for experimentally relevant timescales.
This hints towards the existence of a generalized Gibbs ensemble~\cite{LangenRev, Schmiedmayer_Gibbs} describing the NP and SP. 
In the D-NP, the system tends to heat up. The heating across the entire phase is illustrated by the coloured tiling of the phase diagrams in Fig.~\ref{PDs}.
Remarkably, the D-NP obtained from the NP (upper panel of Fig.~\ref{PDs}) has minimal heating as compared to the D-NP obtained from the SP; it displays pre-thermalization-like plateaus 
where the average energy is approximately constant [see thick solid curve in Fig.~\ref{energy-profiles}(c)] and $\Theta$ shows smooth oscillatory behavior. 
As the amplitude is gradually increased, the width of the plateaus shrinks and the condensate thermalizes more quickly. This rather stable behavior of the D-NP makes it easy to observe experimentally. 
The time scale over which the system absorbs energy depends crucially on the static pump rate $\eta_p^0$, the amplitude $\alpha$ and the period $T$.

\begin{figure}[t]
\includegraphics[width=\columnwidth]{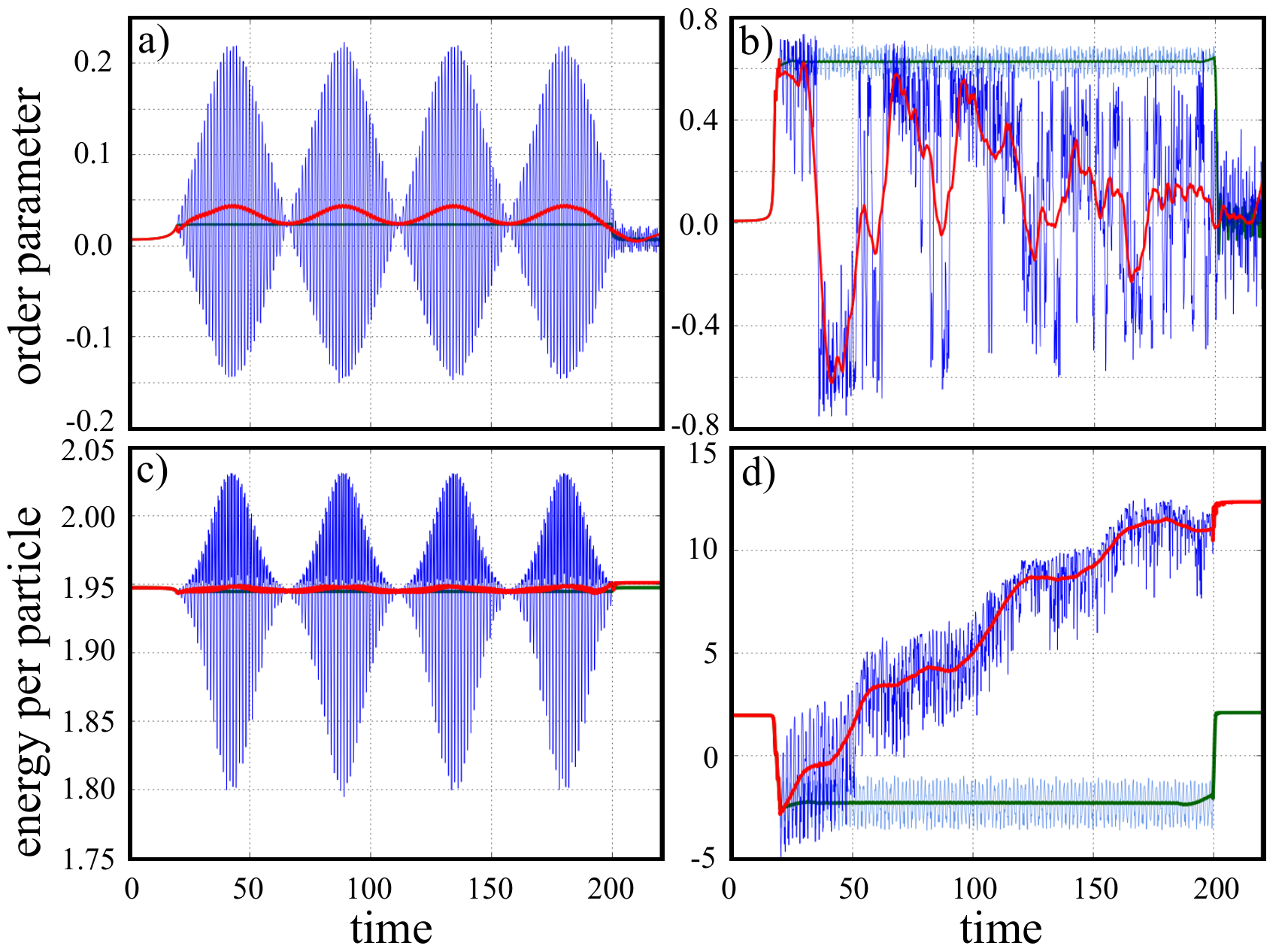}
\caption{Time evolution of the order parameter and energy for a system driven from the NP (left panels) and SP (right panels). Thin lines represent the raw data, while running averages are marked 
as thick solid lines. Each panel features one time-evolution in the modulated NP or SP and another one in the corresponding D-NP at higher periods. 
The driving parameters $\alpha$ and $T$ are $(0.005, 1.6)$ (light line) and $(0.05, 1.35)$ (dark line) for the left panels, and  $(0.04, 1.6)$ (light line) and $(0.096, 1.6)$ (dark line) for the right panels.
In the driven NP we note that the amplitude of the oscillations in the raw energy curve is very small and hence they are covered by their running average (thick dark line). 
All quantities shown are dimensionless, see~\cite{supmat}. }
\label{energy-profiles}
\end{figure}

When obtained from the SP, the D-NP, however, shows a fast thermalization to a trivial high temperature state [see Fig.~\ref{energy-profiles}(d)].  
The corresponding $\Theta$ shows the even to odd lattice reconfigurations, but is increasingly noisy.
The jump from the non-heating (SP) to the heating behavior (D-NP) is sharp, suggesting a first-order phase transition. 
We find that the energy in the D-NP averaged over a period initially increases linearly with time despite the cavity dissipation. 
The heating in the D-NP also signals the breakdown of the mapping to the DM as the system populates other momentum states beyond the integrable subspace of the $ \pm k$ momentum states, 
cf.~\cite{PRLAxel}.
Dissipation shifts the phase boundaries of the D-NP analogously to parametric oscillators and affects the thermalization rate, but it does not change the qualitative features of the system within that phase. 
To summarize, heating profiles of the dissipative BEC-cavity  system  show  an intriguing parametrically-induced crossover between  an effective integrability (where the system does not heat
up)  and non-integrability (where the system heats up).


\textit{Conclusions} -- 
We have investigated the full time-evolution of a BEC coupled to a dissipative high-finesse optical cavity subjected to a time-dependent transverse pumping laser power. 
We show that in addition to the static normal and superradiant phases, 
parametric instabilities  lead to the formation of a new phase (D-NP) where the atoms switch quasiperiodically between the  even- and odd-symmetric configurations. 
Such oscillations are explicitly forbidden in the undriven system. The boundaries of this dynamical normal phase are delineated by Arnold instability lobes. 
The driven NP and SP are resistant to heating -- possibly pointing towards the existence of a generalized Gibbs ensemble. 
The D-NP phase, instead, shows pre-thermalization and eventually thermalizes fully despite the presence of dissipation. 
Possible future directions of research include  investigations of  the nature of the D-NP phase transition, its heating characteristics, analyses of the correlation functions within the condensate, 
probing the effect of fluctuations beyond mean-field and the inclusion of additional optical potential landscapes to compete with the D-NP.
An important outlook would be the generalization of such switching phases to other symmetry classes.


\acknowledgements{The authors gratefully acknowledge C. Bruder, T. Donner and O. Zilberberg for helpful discussions and comments on our work. 
We acknowledge financial support from the Swiss National Science Foundation and Mr. G. Anderheggen.
A. U. J. L. acknowledges financial support by the Swiss SNF and the NCCR Quantum Science and Technology and 
by the Austrian Science Foundation (FWF) under grant No. F65 (SFB ``Complexity in PDEs'') and the Wiener
Wissenschafts- und TechnologieFonds (WWTF) project  No MA16-066 (``SEQUEX''). Computation time on the Hazel Hen cluster of the HLRS in Stuttgart is gratefully acknowledged.}



\begin{thebibliography}{99}







\bibitem{Ritter2009} S. Ritter, F. Brennecke, K. Baumann, T. Donner, C. Guerlin, and T. Esslinger, Appl. Phys. B \textbf{95}, 213 (2009).
\bibitem{PRLPurdy} T. P. Purdy, D. W. C. Brooks, T. Botter, N. Brahms, Z.-Y. Ma, and D. M. Stamper-Kurn, Phys. Rev. Lett. \textbf{105}, 133602 (2010).
\bibitem{eckardt} A. Eckardt and E. Anisimovas, New Journal of Physics \textbf{17}, 093039 (2015).
\bibitem{PRL_Meinert2016} F. Meinert, M. J. Mark, K. Lauber, A. J. Daley, and H.-C. N\"{a}gerl, Phys. Rev. Lett. \textbf{116}, 205301 (2016).
\bibitem{goldmanPRX} N. Goldman and J. Dalibard, Phys. Rev. X \textbf{4}, 031027 (2014).
\bibitem{Landau_Lifshitz} L. Landau and E. Lifshitz, \textit{Mechanics}, Butterworth-Heinemann (1976).
\bibitem{Casati} G. Casati, B. V. Chirikov, F. M. Izraelev, and J. Ford,
\textit{Stochastic Behavior in Classical and Quantum Hamiltonian Systems} (Springer, Berlin, 1979), vol. 93 of Lecture Notes in Physics, pp. 334-352.
\bibitem{Fishman} S. Fishman, D. R. Grempel, and R. E. Prange, Phys. Rev. Lett. \textbf{49}, 509 (1982).
\bibitem{Lazarides} A. Lazarides, A. Das, and R. Moessner, Phys. Rev. E \textbf{90}, 012110 (2014).
\bibitem{Alessio} L. D'Alessio and M. Rigol, Phys. Rev. X \textbf{4}, 041048 (2014).
\bibitem{Ponte} P. Ponte, A. Chandran, Z. Papic, and D. A. Abanin, Annals of Physics \textbf{353}, 196 (2015).
\bibitem{Keeling} J. Keeling, L. M. Sieberer, E. Altman, L. Chen, S. Diehl, and J. Toner, ArXiv (2016),1601.04495.
\bibitem{Dicke:1954} R. H. Dicke, Phys. Rev. \textbf{93}, p. 99 (1954).
\bibitem{Hepp:1973} K. Hepp and E. Lieb, Physical Review \textbf{8}, 2517 (1973).
\bibitem{PRLdomokos} P. Domokos and H. Ritsch, Phys. Rev. Lett. \textbf{89}, 253003 (2002).
\bibitem{Dimer} F. Dimer, B. Estienne, A. S. Parkins, and H. J. Carmichael, Phys. Rev. A \textbf{75}, 013804 (2007).
\bibitem{Baumann:2010} K. Baumann, C. Guerlin, F. Brennecke, and T. Esslinger, Nature \textbf{464}, 1301 (2010).
\bibitem{Baumann:2011}  K. Baumann, R. Mottl, F. Brennecke, and T. Esslinger, Phys. Rev. Lett. \textbf{107}, 140402 (2011).
\bibitem{Klinder17032015} J. Klinder, H. Kessler, M.Wolke, L. Mathey, and A. Hemmerich, Proceedings of the National Academy of Sciences \textbf{112}, 3290 (2015).
\bibitem{PRLBakhtiari}  M. R. Bakhtiari, A. Hemmerich, H. Ritsch, and M. Thorwart, Phys. Rev. Lett. \textbf{114}, 123601 (2015).
\bibitem{Brennecke:2013} F. Brennecke, R. Mottl, K. Baumann, R. Landig, T. Donner, and T. Esslinger, Proc. Natl. Acad. Sci. U.S.A. \textbf{110 (29)}, 11763 (2013).
\bibitem{bastidasPRL} V. M. Bastidas, C. Emary, B. Regler, and T. Brandes, Phys. Rev. Lett. \textbf{108}, 043003 (2012).
\bibitem{Chitra:2015} R. Chitra and O. Zilberberg, Physical Review A \textbf{92}, 023815 (2015).
\bibitem{Nagy:2008} D. Nagy, G. Szirmai, and P. Domokos, Eur. Phys. J. D \textbf{48}, 127 (2008).
\bibitem{HuangSM} K. Huang, \textit{Statistical Mechanics} (Wiley, New York,1987).
\bibitem{Olshanii}  M. Olshanii, Phys. Rev. Lett. 81, \textbf{938} (1998).
\bibitem{PethickBEC} C. J. Pethick and H. Smith, \textit{Bose-Einstein Condensation in Dilute Gases} (Cambridge University Press, Cambridge, 2008), 2nd ed.
\bibitem{PRLNagy2010} D. Nagy, G. K\'{o}nya, G. Szirmai, and P. Domokos, Phys. Rev. Lett. \textbf{104}, 130401 (2010).
\bibitem{supmat} For additional details, see Supplemental Material.
\bibitem{ultracold} A. U. J. Lode, M. C. Tsatsos, and E. Fasshauer, The time-dependent multiconfigurational Hartree method for indistinguishable particles software, URL \url{http://ultracold.org}.
\bibitem{exact_F} E. Fasshauer and A. U. J. Lode, Phys. Rev. A \textbf{93}, 033635 (2016).
\bibitem{SMCTDHB} A. U. J. Lode, Phys. Rev. A \textbf{93}, 063601 (2016).
\bibitem{MCTDHX} O. E. Alon, A. I. Streltsov, and L. S. Cederbaum, J. Chem. Phys. \textbf{127}, 154103 (2007).
\bibitem{PRLAxel} A. U. J. Lode and C. Bruder, Phys. Rev. Lett. \textbf{118}, 013603 (2017).
\bibitem{MCTDHB_init} O. E. Alon, A. I. Streltsov, and L. S. Cederbaum, Phys. Rev. A \textbf{77}, 033613 (2008).
\bibitem{MCTDHB_PRL} A. I. Streltsov, O. E. Alon, and L. S. Cederbaum, Phys. Rev. Lett. \textbf{99}, 030402 (2007).
\bibitem{Emary} C. Emary and T. Brandes, Phys. Rev. E \textbf{67}, 066203 (2003).
\bibitem{HillsEq} W. Magnus and S. Winkler, \textit{Hill's equation} (Interscience Publishers, 1966).
\bibitem{mclachlan} N. McLachlan, \textit{Theory and application of Mathieu functions} (Clarendon, 1951).
\bibitem{LangenRev}  T. Langen, T. Gasenzer, and J. Schmiedmayer, J. Stat. Mech.: Theor. Exp., p. 064009 (2016).
\bibitem{Schmiedmayer_Gibbs} T. Langen, S. Erne, R. Geiger, B. Rauer, T. Schweigler, M. Kuhnert, W. Rohringer, I. E. Mazets, T. Gasenzer, and J. Schmiedmayer, Science \textbf{348}, 207 (2015).
\end{thebibliography}
\end{document}